\documentclass[preprint,12pt]{elsarticle}

\usepackage{amssymb}
\usepackage{amsmath}
\usepackage{graphicx}
\usepackage{booktabs}
\usepackage{url}
\usepackage{listings}
\usepackage{microtype}
\usepackage{placeins}

\newcounter{bla}

\journal{Computer Physics Communications}

\begin{document}

\begin{frontmatter}

\title{symveig: Verified eigenvalue enclosures for symmetry-decomposed
Hermitian matrices}

\author[a]{Sarang Vehale\corref{author}}
\ead{sarangvehale2@gmail.com}
\author[b]{Ritu Goel}

\cortext[author]{Corresponding author.}

\address[a]{National Forensic Sciences University, Delhi, India}
\address[b]{Department of Applied Science and Humanities, School of
Engineering \& Technology, Vivekananda Institute of Professional Studies
- Technical Campus, AU-Block (Outer Ring Road), Pitampura, Delhi -
110034, India}

\begin{abstract}
Exact diagonalization of quantum lattice Hamiltonians returns
floating-point eigenvalues whose accuracy is not certified: rounding
error, eigensolver behaviour, and ill-conditioning can corrupt a result
without warning. We present \texttt{symveig}, a pure NumPy/SciPy package
that computes rigorous, machine-checkable enclosures of all eigenvalues
of a Hermitian matrix, with an optional symmetry-sector decomposition.
For a matrix that commutes with an abelian conserved quantity diagonal in
the working basis (for example the total magnetization of a spin model),
the package verifies each symmetry sector independently. Because each
sector block is much smaller than the full matrix, this yields enclosures
that are both tighter (by a factor of $3$--$9$ across system sizes
$L = 4$--$12$) and dramatically faster (a wall-clock speedup of up to
$130\times$ at $L = 12$) than verifying the full matrix, while never
forming or diagonalizing it. Every enclosure half-width is a guaranteed
upper bound on the distance from a computed eigenvalue to the nearest
true eigenvalue under IEEE~754 round-to-nearest arithmetic, obtained by
explicit floating-point error analysis with no heuristic slack. The
implementation requires neither INTLAB nor MATLAB, bringing rigorously
verified eigenvalue enclosure into the standard scientific-Python stack
used in computational physics. We validate the package on
$1$D Heisenberg (open and periodic), $J_1$--$J_2$ Heisenberg, and $2$D
Heisenberg lattices, confirming that every computed eigenvalue is
contained in its enclosure across all tested configurations.

\noindent \textbf{PROGRAM SUMMARY}

\begin{small}
\noindent
{\em Program Title:} symveig                                  \\
{\em CPC Library link to program files:} (to be added by Technical Editor) \\
{\em Developer's repository link:} \url{https://github.com/sarang-kernel/symveig.git} \\
{\em Zenodo Archive:} \url{https://doi.org/10.5281/zenodo.20695554}
{\em Licensing provisions:} MIT  \\
{\em Programming language:} Python 3 ($\geq 3.9$)             \\
{\em Supplementary material:} Frozen example outputs (JSON results,
summary CSV, and figures) reproducing the tables and figures of this
paper are included in the \texttt{results/} directory of the archived
release. \\
{\em Nature of problem:}\\
Exact diagonalization of quantum lattice Hamiltonians and other Hermitian
operators returns floating-point eigenvalues whose accuracy is not
certified. Rounding error, eigensolver convergence behaviour, and
ill-conditioning can all corrupt the result without warning. For
applications that require guaranteed bounds (confirming a spectral
gap, certifying a ground-state energy, or validating an approximate
solver), one needs a rigorous interval proven to contain the true
eigenvalue. Existing verified eigensolvers of this kind require MATLAB
and the commercial INTLAB toolbox, which are not part of the typical
computational-physics software stack and do not natively exploit the
symmetry-sector structure that quantum lattice models possess. \\
{\em Solution method:}\\
For a Hermitian matrix $A$, \texttt{symveig} computes an approximate
eigendecomposition with LAPACK (via NumPy) and bounds the distance from
each computed eigenvalue to the nearest true eigenvalue using the
Hermitian Bauer--Fike--Wilkinson residual estimate. Every floating-point
operation entering the residual and norm computations is bounded
rigorously with Higham's $\gamma_k$ error analysis, so the returned
interval is a guaranteed enclosure under IEEE~754 round-to-nearest
arithmetic. Overlapping per-eigenvalue intervals are merged into
rigorous multiplicity-counted cluster enclosures. When the operator
commutes with an abelian conserved quantity diagonal in the working
basis, the package projects $A$ into each symmetry sector and verifies
the sectors independently; because each sector block has dimension at
most $\sim \binom{L}{L/2}$, this yields tighter enclosures and a large
wall-clock speedup relative to verifying the full matrix. \\
{\em Additional comments including restrictions and unusual features:}\\
The package is pure NumPy/SciPy; no INTLAB, MATLAB, or compiled
extensions are required. The dominant cost is the dense
eigendecomposition, $O(n^3)$; the sector path avoids ever diagonalizing
the full matrix and is the recommended mode of use. Dense methods are
practical to dimension $\sim 4096$ ($1$D spin-$1/2$ chains up to
$L = 12$) on a workstation. An optional cluster-refinement tier is
provided but disabled by default, as without directed-rounding
arithmetic it is dominated by worst-case error bounds. The verification
assumes the symmetry observable is exactly diagonal in the working basis,
which holds for the standard abelian $U(1)$-type conserved quantities of
lattice models. \\

\end{small}
\end{abstract}

\end{frontmatter}

\section{Introduction}
\label{sec:intro}

Exact diagonalization (ED) is a workhorse of computational quantum
many-body physics: for a lattice Hamiltonian small enough to be
represented as a dense or sparse matrix, ED returns the spectrum and
eigenstates to within the accuracy of the underlying linear-algebra
routines~\cite{Sandvik2010,QuSpin2017,XDiag2025}. The results are
trusted implicitly. In the overwhelming majority of applications this
trust is justified: LAPACK's symmetric eigensolvers~\cite{LAPACK1999}
are backward stable, and the computed eigenvalues are typically accurate
to a relative error of order the unit roundoff.

There is, however, a class of questions for which ``typically accurate''
is not enough. Does a given Hamiltonian have a spectral gap above its
ground state, and if so, is the gap rigorously bounded away from zero?
Is a computed ground-state energy a certified upper or lower bound? Does
an approximate or accelerated eigensolver (a kernel-polynomial
expansion~\cite{Weisse2006KPM}, a Lanczos run with limited
reorthogonalization~\cite{Lanczos1950}, or a machine-learned surrogate)
actually return what it claims, to a guaranteed tolerance? For such
questions one needs not a floating-point number but a \emph{verified
enclosure}: an interval proven to contain the true eigenvalue,
accounting for every rounding error incurred in computing the bound
itself.

The mathematics of verified eigenvalue enclosure is mature. Building on
the perturbation theory of Bauer and Fike~\cite{BauerFike1960} and the
verification framework of Rump~\cite{Rump2010Acta}, a sequence of works
has produced rigorous, $O(n^3)$ algorithms for enclosing all eigenvalues
and invariant subspaces of general and Hermitian matrices, with
automatic detection of eigenvalue
clusters~\cite{Rump2001,Miyajima2010,Miyajima2014,Rump2022,RumpLange2023}.
These methods are implemented in INTLAB~\cite{Rump1999INTLAB}, the de
facto standard interval-arithmetic library. INTLAB is, however, a
commercial MATLAB toolbox. It is not part of the scientific-Python stack
(NumPy~\cite{NumPy2020}, SciPy~\cite{SciPy2020}) on which modern
ED packages such as QuSpin~\cite{QuSpin2017,QuSpin2019},
XDiag~\cite{XDiag2025}, and ALPS~\cite{ALPS2011} are built, and a
physicist who has just diagonalized a Hamiltonian in Python has no
in-ecosystem route to certifying the result.

A second gap is structural. Quantum lattice Hamiltonians are rarely
generic matrices: they almost always commute with one or more conserved
quantities. The simplest and most ubiquitous is an abelian $U(1)$
symmetry (total magnetization $S_z$ for a spin model, total particle
number for a fermionic or bosonic model), which is diagonal in the
computational (occupation) basis and block-diagonalizes the Hamiltonian
into symmetry sectors. ED practitioners exploit this routinely to reduce
memory and time. The general-purpose verified eigensolvers, by contrast,
discover near-degenerate clusters numerically from the spectrum of the
matrix itself; they do not take an externally supplied commuting
observable as an input, and so do not benefit from the sector structure
that is handed to them for free by the physics.

This paper presents \texttt{symveig}, a package that closes both gaps. It
computes rigorous eigenvalue enclosures for Hermitian matrices in pure
NumPy/SciPy, with no dependence on INTLAB or MATLAB, and it accepts an
abelian conserved quantity as an explicit input, verifying each symmetry
sector independently. It complements CERTIFY-ED~\cite{CertifyED2026}, a
recent multi-layer verification framework for exact diagonalization in
the same scientific-Python setting: where CERTIFY-ED certifies an ED
result through agreement of independent oracles, conservation laws, and
arbitrary-precision reference computation, \texttt{symveig} provides the
complementary capability of a single rigorous interval enclosure with a
machine-checkable error bound, reusing the same family of lattice models
for validation. The sector decomposition is advantageous in two
distinct ways that we quantify below: the enclosures are
\emph{tighter}, because the half-width of the rigorous bound scales with
the dimension of the matrix being verified and each sector block is much
smaller than the full matrix; and the computation is \emph{faster},
because the sector path solves several small eigenproblems instead of
one large one and never diagonalizes the full matrix.

We emphasize at the outset what is and is not novel here. The
per-eigenvalue and per-cluster verification mathematics is standard, due
to Rump, Lange, Miyajima, and
others~\cite{Rump2001,Miyajima2014,Rump2022,RumpLange2023}; we
re-implement the Hermitian residual bound with explicit floating-point
error analysis. The contribution of \texttt{symveig} is as a
\emph{program}: an open, dependency-light, well-tested implementation
that makes verified enclosure available in the Python ED ecosystem, that
exploits the abelian symmetry structure of lattice models to produce
tighter and faster certificates, and that we validate empirically on a
family of physically meaningful Hamiltonians. This is, to our knowledge,
the first such tool in the scientific-Python stack.

The remainder of the paper is organized as follows.
Section~\ref{sec:theory} develops the rigorous enclosure bound and the
sector-decomposition argument. Section~\ref{sec:implementation} describes
the software. Section~\ref{sec:results} reports validation and
performance on lattice models. Section~\ref{sec:discussion} discusses
limitations and extensions, and Section~\ref{sec:conclusion} concludes.

\section{Method}
\label{sec:theory}

\subsection{The enclosure problem}

Let $A \in \mathbb{C}^{n \times n}$ be Hermitian. A numerical eigensolver
returns approximate eigenpairs $(\mu_i, x_i)$. We seek, for each $i$, a
rigorous interval $[\mu_i - h_i,\ \mu_i + h_i]$ guaranteed to contain a
true eigenvalue of $A$, where ``guaranteed'' means provably correct
under IEEE~754 double-precision round-to-nearest
arithmetic~\cite{IEEE754_2019}, accounting for every rounding error in
the computation of the bound itself.

\subsection{The Hermitian residual bound}

The foundation is the residual estimate for Hermitian (more generally,
normal) matrices. If $x$ has unit $2$-norm, $\mu = x^{*} A x$, and the
residual is $r = A x - \mu x$, then there exists a true eigenvalue
$\lambda$ of $A$ with
\begin{equation}
\label{eq:bf}
|\lambda - \mu| \le \|r\|_2 .
\end{equation}
This is the Bauer--Fike theorem~\cite{BauerFike1960} specialized to the
Hermitian case, where the eigenvector-matrix condition number is one;
see Parlett~\cite[Thm.~11.7.1]{Parlett1998} or Golub and Van
Loan~\cite{GolubVanLoan2013}. For a computed eigenvector that is only
approximately of unit norm, the bound is $|\lambda - \mu| \le \|r\|_2 /
\|x\|_2$.

\subsection{Making the bound rigorous in floating point}

The quantities $\|r\|_2$ and $\|x\|_2$ are themselves computed in
floating point. To obtain a \emph{provable} upper bound on
$|\lambda - \mu|$ we replace them by a rigorous upper bound on $\|r\|_2$
and a rigorous lower bound on $\|x\|_2$:
\begin{equation}
\label{eq:rigorous}
|\lambda - \mu| \le
\frac{\|r\|_2^{\mathrm{fp}} + \delta_r}
     {\|x\|_2^{\mathrm{fp}} - \delta_x} .
\end{equation}
The error terms $\delta_r$ and $\delta_x$ are bounded using Higham's
$\gamma_k$ notation~\cite{Higham2002}. With unit roundoff
$u = 2^{-53}$ for IEEE~754 double precision,
\begin{equation}
\gamma_k = \frac{k\,u}{1 - k\,u}, \qquad k\,u < 1 ,
\end{equation}
and for complex arithmetic we replace $u$ by $\sqrt{2}\,u$, absorbing the
cost of complex multiply--add into the constant~\cite[Sec.~3.6]{Higham2002}.
The relevant contributions are:
\begin{itemize}
\item Matrix--vector product $A x$: each entry is a length-$n$ inner
  product, so $\|\mathrm{fl}(Ax) - Ax\|_2 \le \gamma_{4n}\,\|A\|_F\,\|x\|_2$.
\item Scalar--vector product $\mu x$: error $\le \gamma_1\,|\mu|\,\|x\|_2$.
\item Subtraction $Ax - \mu x$: error
  $\le \gamma_1(\|Ax\| + \|\mu x\|)$.
\item The $2$-norm: relative error $\le \gamma_{2n}$.
\item The Frobenius norm $\|A\|_F$: computed with relative error
  $\le \gamma_{2n^2}$, inflated so that it enters the bound as a rigorous
  upper bound.
\end{itemize}
Combining these gives a closed-form, rigorously valid $\delta_r$, and the
lower bound $\|x\|_2 \ge \|x\|_2^{\mathrm{fp}} / (1 + \gamma_{2n})$ gives
$\delta_x$. The resulting half-width
\begin{equation}
\label{eq:halfwidth}
h_i = \frac{\|r_i\|_2^{\mathrm{fp}} + \delta_{r,i}}
           {\|x_i\|_2^{\mathrm{fp}} / (1 + \gamma_{2n})}
\end{equation}
is a guaranteed bound on the distance from $\mu_i$ to the nearest true
eigenvalue. There is no heuristic slack; every term is derived from the
floating-point model.

In practice the bound is dominated not by the true residual, which is
typically of order $n\,u\,\|A\|$ from a backward-stable eigensolver,
but by the worst-case constant $\gamma_{4n}\|A\|_F$ in $\delta_r$. This
is the price of full rigour without directed-rounding arithmetic. As we
show empirically in Section~\ref{sec:results}, the half-width
consequently scales as $h \approx C\,u\,d\,\|A\|_F$ with a
model-independent constant $C$ of order a few, where $d$ is the matrix
dimension.

\subsection{Clusters and multiplicities}

Hermitian lattice models routinely have exact degeneracies imposed by
symmetry, so several of the intervals from~\eqref{eq:halfwidth} overlap.
We merge any maximal chain of overlapping intervals into a single
\emph{cluster enclosure} $[c_{\mathrm{lo}}, c_{\mathrm{hi}}]$ and report
it with multiplicity equal to the number of merged eigenvalues. The
guarantee upgrades accordingly: the interval contains exactly that many
true eigenvalues, counted with algebraic multiplicity. Because the merge
is a rigorous interval union, containment is preserved.

\subsection{Symmetry-sector decomposition}
\label{sec:sectors}

Suppose $A$ commutes with a Hermitian observable $S$, $[A, S] = 0$, and
$S$ is diagonal in the working basis with sectors (distinct eigenvalues)
$s$. Then $A$ is block diagonal in that basis: it preserves each
eigenspace of $S$. Writing $P_s$ for the projector onto sector $s$, the
sector block $A_s = P_s A P_s$ is obtained simply by selecting the rows
and columns whose $S$-value equals $s$; no numerical diagonalization of
$S$ is required, since $S$ is already diagonal. Each $A_s$ is Hermitian
of dimension $d_s$, and the spectrum of $A$ is the disjoint union of the
spectra of the $A_s$. Verifying each block independently and taking the
union of the enclosures therefore reconstructs a verified enclosure of
the full spectrum.

The benefit is twofold, and both parts follow from the scaling
$h \approx C\,u\,d\,\|A\|_F$:
\begin{enumerate}
\item \textbf{Tighter enclosures.} The worst sector block has dimension
  $\max_s d_s$, which for a $1$D spin-$1/2$ chain is
  $\binom{L}{L/2} \approx n/\sqrt{L}$, roughly $n/4$ at accessible sizes.
  Since the half-width scales with the dimension, the worst sector
  half-width is smaller than the global half-width by approximately
  $n / \max_s d_s$, modulated by the ratio $\|A\|_F / \|A_s\|_F$.
\item \textbf{Faster computation.} The global path performs one $O(n^3)$
  eigendecomposition; the sector path performs several $O(d_s^3)$
  eigendecompositions with $\sum_s d_s = n$ and $\sum_s d_s^3 \ll n^3$,
  and never diagonalizes the full matrix.
\end{enumerate}
The sector path is therefore the recommended mode of use; the global
path exists in our benchmark only to provide the comparison baseline.

\section{Software description}
\label{sec:implementation}

\subsection{Overview and dependencies}

\texttt{symveig} is a Python~3 package ($\geq 3.9$) depending only on
NumPy~\cite{NumPy2020} and SciPy~\cite{SciPy2020}; Matplotlib is an
optional dependency used only for figure generation. It is installable
with \texttt{pip install .} and exposes both a library API and a
one-command reproducibility driver.

\subsection{Public interface}

The core API consists of two functions and a result type. For a
Hermitian array \texttt{A},
\begin{lstlisting}
from symveig import enclose_global, enclose_sectors
encs = enclose_global(A)             # verified enclosures, all eigenvalues
encs = enclose_sectors(A, S_diag)    # sector-decomposed enclosures
\end{lstlisting}
Both return a list of \texttt{Enclosure} objects, each carrying a
midpoint, a rigorous half-width, an integer multiplicity, and the name
of the bound that produced it. The interval $[\,$\texttt{lo}$,\,$%
\texttt{hi}$\,]$ of each \texttt{Enclosure} is guaranteed to contain
exactly \texttt{multiplicity} true eigenvalues, and the union over all
returned objects contains the full spectrum.

\subsection{Module structure}

The package is organized as follows. \texttt{certify\_eig.py} implements
the enclosure engine: the per-eigenvalue bound~\eqref{eq:halfwidth}, the
cluster merge, and an optional refinement tier (Section~\ref{sec:tier2}).
\texttt{models.py} provides builders for the lattice Hamiltonians used in
testing and benchmarking ($1$D Heisenberg with open or periodic
boundaries, transverse-field Ising, $J_1$--$J_2$ Heisenberg, and $2$D
Heisenberg), each returning the Hamiltonian together with the diagonal of
its conserved $S_z$ where applicable. \texttt{benchmark.py} drives a
single model, computing global and sector enclosures, the width ratio,
and the wall-clock speedup. \texttt{plot\_results.py} renders the figures
from saved JSON. \texttt{cli.py} orchestrates the full reproducibility
run, and a thin \texttt{run.py} wrapper exposes it from the repository
root.

\subsection{Reproducibility driver}

A single command,
\begin{center}
\begin{minipage}{0.5\linewidth}
\begin{verbatim}
`python run.py`
\end{verbatim}
\end{minipage}
\end{center}
runs the unit-test suite, the benchmark over all models, and the figure
generation, writing all outputs to a \texttt{results/} directory:
platform and version metadata, the test log, one JSON file per model, a
summary CSV, and the figures as PDF and PNG. The flag \texttt{--full}
extends the benchmark to the largest sizes ($L = 12$, dimension $4096$,
and a $4 \times 3$ two-dimensional lattice); because the global baseline
at that size performs a dense complex eigendecomposition that saturates
all cores, \texttt{--full} auto-throttles the BLAS thread count, and a
\texttt{--threads} option gives explicit control.

\subsection{Optional cluster-refinement tier}
\label{sec:tier2}

The package includes a second-tier cluster bound following the
Davis--Kahan~\cite{DavisKahan1970} analysis of Rump and
Lange~\cite[Lemma~6.1]{RumpLange2023}, which bounds the deviation of the
true cluster eigenvalues from the cluster centroid using the spectral
norm of the block residual and the spectral gap to the rest of the
spectrum. In the present release this tier is \emph{disabled by default}:
without directed-rounding arithmetic, the rigorous floating-point bound
on the block residual inherits the same worst-case
$\gamma_{4n}\|A\|_F$ term as the first-tier bound, plus a factor that
grows with cluster size, so the refined bound is typically looser than
the merged first-tier interval. We retain it as an opt-in feature and
document the limitation; tightening it with directed rounding is
identified as future work (Section~\ref{sec:discussion}).

\subsection{Testing}

The test suite covers five classes of synthetic matrices, namely isolated
spectra, a constructed high-multiplicity cluster, a block-structured
matrix with a known sector partition, an ill-conditioned matrix with
condition number $\sim 10^{16}$, and a near-degenerate pair separated by
$10^{-10}$, checking in each case that every reference eigenvalue lies
in some returned enclosure. Model tests verify Hermiticity, the
commutation $[A, S] = 0$, and the combinatorial sector sizes
$\binom{L}{L/2 + s}$. A determinism test confirms that repeated runs on
the same machine produce bit-identical output.

\section{Results and validation}
\label{sec:results}

We validate \texttt{symveig} on four families of Hermitian lattice
Hamiltonians drawn from the model library of
CERTIFY-ED~\cite{CertifyED2026}: the $1$D spin-$1/2$ Heisenberg chain
with open (OBC) and periodic (PBC) boundaries, the $1$D $J_1$--$J_2$
Heisenberg chain at $J_2 = 0.5$ (PBC), and the $2$D Heisenberg model on
$3 \times 3$ and $4 \times 3$ lattices (PBC). All four conserve the total
magnetization $S_z$, which is diagonal in the computational basis and
supplies the sector decomposition. All results below were produced by a
single \texttt{python run.py --full} invocation on a Linux x86-64
workstation (Python~3.14, NumPy~2.4); the complete output is included
with the package.

\subsection{Rigour: containment}

The primary correctness criterion for a verified solver is that the true
eigenvalues actually lie inside the returned intervals. For every one of
the $17$ (model, $L$) configurations tested, and for both the global and
the sector paths, every reference eigenvalue computed by an independent
LAPACK call is contained in some returned enclosure, with the
multiplicities summing to the matrix dimension. The package thus returns
sound certificates across the full range of sizes, boundary conditions,
and lattice geometries tested, including the deliberately ill-conditioned
and near-degenerate synthetic cases of the test suite.

\subsection{Enclosure tightness and its scaling}

Table~\ref{tab:main} reports, for each configuration, the maximum
half-width of the global and sector enclosures and their ratio. The
sector enclosures are uniformly tighter, by a factor that grows smoothly
from $\approx 3.2$ at $L = 4$ to $\approx 9.1$ at $L = 12$, and the
pattern is essentially identical across all four model families and both
lattice dimensionalities.

\begin{table}[t]
\centering
\caption{Maximum enclosure half-width for the global and sector paths,
and their ratio, across all tested configurations. Half-widths are
rigorous bounds on $|\lambda_{\mathrm{true}} - \mu|$. Every reference
eigenvalue is contained in both the global and the sector enclosures in
every row.}
\vspace{0.4cm}
\label{tab:main}
\small
\begin{tabular}{l r r r r r}
\toprule
Model & $L$ & $\dim$ & global $h_{\max}$ & sector $h_{\max}$ & ratio \\
\midrule
1D Heisenberg OBC      &  4 &   16 & $3.15\times10^{-14}$ & $9.33\times10^{-15}$ & 3.38 \\
1D Heisenberg OBC      &  6 &   64 & $3.16\times10^{-13}$ & $6.31\times10^{-14}$ & 5.01 \\
1D Heisenberg OBC      &  8 &  256 & $2.96\times10^{-12}$ & $4.56\times10^{-13}$ & 6.49 \\
1D Heisenberg OBC      & 10 & 1024 & $2.68\times10^{-11}$ & $3.44\times10^{-12}$ & 7.77 \\
1D Heisenberg OBC      & 12 & 4096 & $2.36\times10^{-10}$ & $2.64\times10^{-11}$ & 8.94 \\
1D Heisenberg PBC      &  4 &   16 & $3.73\times10^{-14}$ & $1.17\times10^{-14}$ & 3.20 \\
1D Heisenberg PBC      &  6 &   64 & $3.46\times10^{-13}$ & $6.93\times10^{-14}$ & 4.99 \\
1D Heisenberg PBC      &  8 &  256 & $3.16\times10^{-12}$ & $4.88\times10^{-13}$ & 6.47 \\
1D Heisenberg PBC      & 10 & 1024 & $2.82\times10^{-11}$ & $3.64\times10^{-12}$ & 7.75 \\
1D Heisenberg PBC      & 12 & 4096 & $2.47\times10^{-10}$ & $2.77\times10^{-11}$ & 8.93 \\
1D $J_1$--$J_2$ PBC    &  4 &   16 & $4.46\times10^{-14}$ & $1.20\times10^{-14}$ & 3.72 \\
1D $J_1$--$J_2$ PBC    &  6 &   64 & $3.87\times10^{-13}$ & $7.42\times10^{-14}$ & 5.22 \\
1D $J_1$--$J_2$ PBC    &  8 &  256 & $3.53\times10^{-12}$ & $5.30\times10^{-13}$ & 6.66 \\
1D $J_1$--$J_2$ PBC    & 10 & 1024 & $3.15\times10^{-11}$ & $3.99\times10^{-12}$ & 7.90 \\
1D $J_1$--$J_2$ PBC    & 12 & 4096 & $2.76\times10^{-10}$ & $3.05\times10^{-11}$ & 9.07 \\
2D Heisenberg $3\times3$ PBC & 9 &  512 & $1.34\times10^{-11}$ & $1.67\times10^{-12}$ & 8.04 \\
2D Heisenberg $4\times3$ PBC & 12 & 4096 & $3.49\times10^{-10}$ & $3.84\times10^{-11}$ & 9.10 \\
\bottomrule
\end{tabular}
\end{table}

\begin{figure}[!htbp]
\centering
\includegraphics[width=0.65\linewidth]{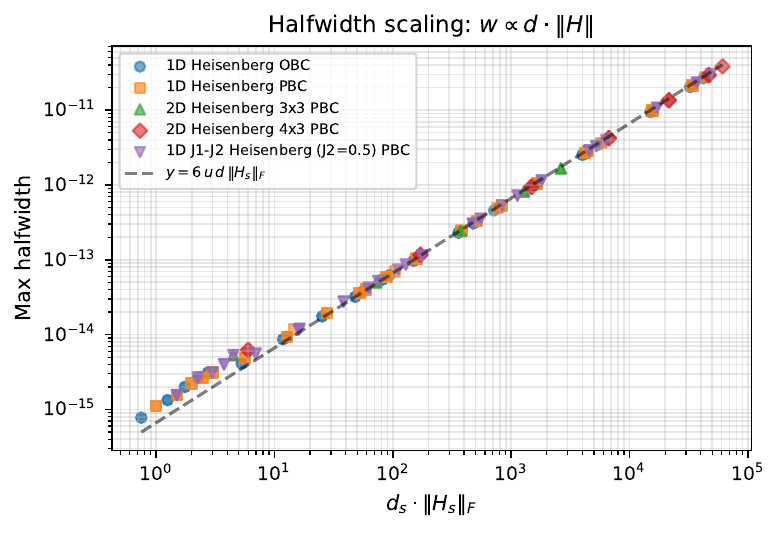}
\caption{Maximum per-sector enclosure half-width versus
$d_s\,\|A_s\|_F$, for every sector of every model and size. The data
collapse onto the line $h \approx 6\,u\,d_s\,\|A_s\|_F$
($u = 2^{-53}$) across more than four orders of magnitude, demonstrating
that the rigorous half-width is a predictable function of sector
dimension and norm.}
\label{fig:scaling}
\end{figure}

\begin{figure}[!htbp]
\centering
\includegraphics[width=0.64\linewidth]{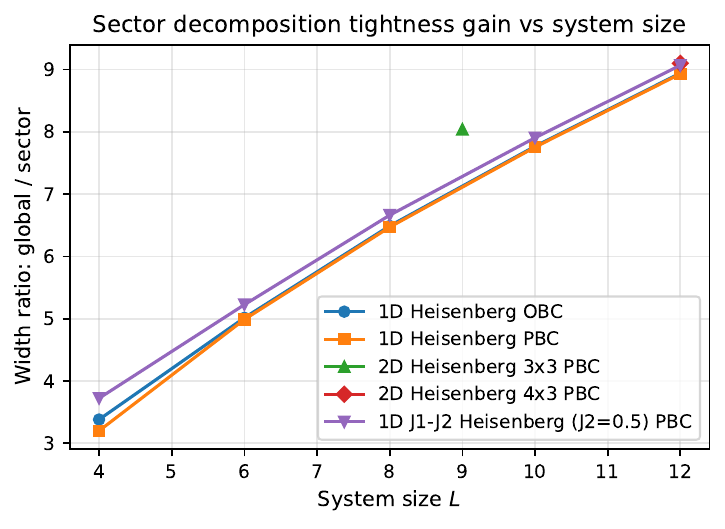}
\caption{Ratio of the maximum global enclosure half-width to the maximum
sector enclosure half-width, as a function of system size $L$, for all
model families. The sector decomposition is uniformly tighter, with the
advantage growing with sy4tem size.}
\label{fig:ratio}
\end{figure}

This behaviour is explained quantitatively by the scaling argument of
Section~\ref{sec:theory}. Figure~\ref{fig:scaling} plots the maximum
per-sector half-width against $d_s\,\|A_s\|_F$ for every sector of every
model. The points collapse onto a single line over more than four orders
of magnitude in $d_s\,\|A_s\|_F$, confirming
$h \approx C\,u\,d\,\|A\|_F$ with $C \approx 6$ and $u = 2^{-53}$,
independent of model family, boundary conditions, and lattice dimension.
The tightness gain of the sector decomposition is thus a direct and
predictable consequence of replacing the dimension $n$ of the full matrix
by the much smaller dimension $\max_s d_s$ of the largest sector block.
Figure~\ref{fig:ratio} shows the resulting global-to-sector width ratio
as a function of $L$.

\FloatBarrier
\subsection{Performance}

\begin{figure}[!htbp]
\centering
\includegraphics[width=0.6\linewidth]{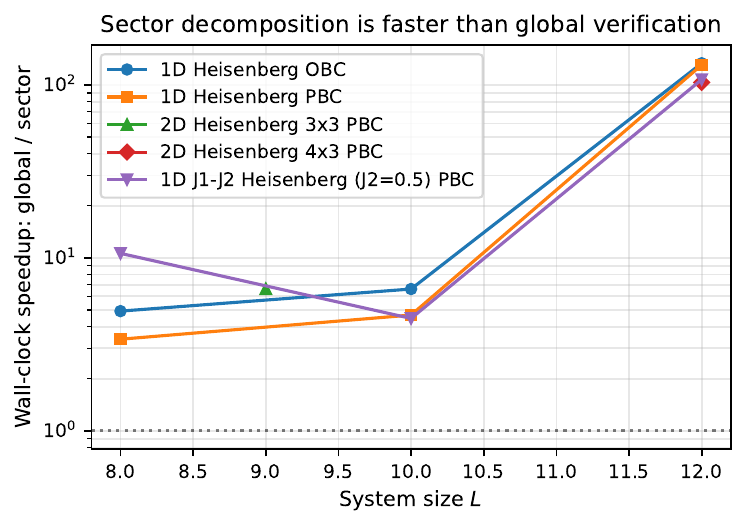}
\caption{Wall-clock speedup of the sector path over the global baseline
versus system size $L$ (logarithmic vertical axis). The dotted line marks
parity. The speedup grows rapidly with size as the cost of the full
dense eigendecomposition comes to dominate the global path.}
\label{fig:speedup}
\end{figure}

Because the sector path never diagonalizes the full matrix, it
is also substantially faster. Table~\ref{tab:timing} and
Figure~\ref{fig:speedup} report the wall-clock speedup of the sector path
over the global baseline. At the largest sizes the global path is
dominated by a single dense complex Hermitian eigendecomposition of
dimension $4096$ (about $56$~s on the test machine), whereas the sector
path solves the symmetry blocks (the largest of dimension
$\binom{12}{6} = 924$) in about half a second, a speedup exceeding
$100\times$. At smaller sizes the speedup is more modest but always
favourable once the matrices are large enough for the cost to be
dominated by linear algebra rather than Python overhead.

\begin{table}[!htbp]
\centering
\caption{Wall-clock time of the global eigendecomposition baseline and
the sector path, and their ratio, for the largest size of each model
family. Times in seconds on a Linux x86-64 workstation.}
\vspace{0.3cm}
\label{tab:timing}
\small
\begin{tabular}{l r r r r}
\toprule
Model & $L$ & global (s) & sector (s) & speedup \\
\midrule
1D Heisenberg OBC          & 12 & 55.5 & 0.455 & 133.5 \\
1D Heisenberg PBC          & 12 & 56.0 & 0.448 & 130.7 \\
1D $J_1$--$J_2$ PBC        & 12 & 55.8 & 0.539 & 107.6 \\
2D Heisenberg $4\times3$   & 12 & 57.0 & 0.576 & 103.5 \\
\bottomrule
\end{tabular}
\end{table}

\subsection{Per-sector structure}

\begin{figure}[!htbp]
\centering
\includegraphics[width=0.8\linewidth]{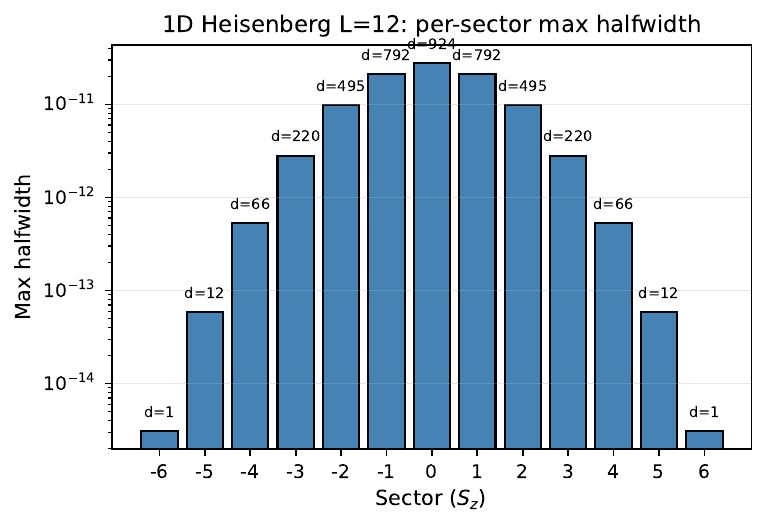}
\caption{Per-sector maximum enclosure half-width for the $1$D Heisenberg
chain at $L = 10$ (logarithmic vertical axis), labelled by sector
dimension $d$. The half-width grows with sector dimension, in agreement
with the scaling law of Figure~\ref{fig:scaling}.}
\label{fig:persector}
\end{figure}

Figure~\ref{fig:persector} shows the per-sector maximum half-width for
the $1$D Heisenberg chain at $L = 10$, with each bar labelled by the
sector dimension. The half-width tracks the sector dimension exactly as
the scaling law predicts: the largest sectors near $S_z = 0$ carry the
widest enclosures, and the singleton sectors at $S_z = \pm L/2$ are
verified to near machine precision. This is the mechanism behind the
overall tightness gain: by verifying each sector at its own (small)
dimension rather than at the full dimension $n$, every eigenvalue
receives the tightest bound its sector allows.

\section{Discussion}
\label{sec:discussion}

\subsection{Relation to existing verified eigensolvers}

The per-eigenvalue and per-cluster verification mathematics implemented
here is standard, originating with Rump and developed by Miyajima, Rump,
and Lange into complete $O(n^3)$ algorithms for all eigenpairs and
invariant subspaces of Hermitian and general
matrices~\cite{Rump2001,Miyajima2014,Rump2022,RumpLange2023}, available
in INTLAB~\cite{Rump1999INTLAB}. \texttt{symveig} does not advance that
theory. Its role is complementary and practical: it brings a rigorous
Hermitian enclosure into the NumPy/SciPy ecosystem with no commercial or
compiled dependencies, and it adds the symmetry-sector decomposition,
which the general-purpose tools do not provide because they detect
clusters from the spectrum rather than accepting an external commuting
observable. We regard \texttt{symveig} as a bridge between the verified-%
computing literature and the computational-physics ED workflow, not as a
competitor to INTLAB on the underlying numerics.

\subsection{Limitations}

Three limitations bound the present scope. First, the first-tier bound is
dominated by the worst-case $\gamma_{4n}\|A\|_F$ matrix--vector error
constant rather than by the true residual, so the enclosures, while
rigorous, are looser than the directed-rounding bounds achievable in
INTLAB by a factor that grows with dimension. The natural remedy
(computing the residual norm with rounding-mode-controlled arithmetic in
the manner of Rump~\cite{Rump2010Acta}) would let the bound track the
true residual $\sim n\,u\,\|A\|$ and, in particular, would make the
optional cluster-refinement tier (Section~\ref{sec:tier2}) advantageous
rather than redundant. Second, the method is dense: it is practical to
dimension $\sim 4096$, i.e.\ $1$D spin-$1/2$ chains to $L = 12$ or
comparable $2$D clusters, on a workstation. Third, and most importantly
for the sector decomposition, the conserved observable must be exactly
diagonal in the working basis. This holds for the standard abelian
$U(1)$-type quantities (total magnetization, particle number) but
not for momentum, point-group, or non-abelian (full $SU(2)$) symmetries,
whose use as a blocking input would require an additional analysis of how
a numerically diagonalized commuting observable propagates error into the
per-sector enclosures. That cross-matrix analysis is a well-defined
direction for future work.

\subsection{Use cases}

Within its scope the tool is directly useful for certifying ground-state
energies and low-lying gaps of lattice models, for validating the output
of approximate or accelerated eigensolvers against a guaranteed
reference, and as a verification layer for ED packages that currently
return uncertified floating-point spectra. The sector path makes such
certification cheap enough (sub-second for matrices where the
uncertified diagonalization itself takes a minute) that it can be run
routinely rather than reserved for special cases.

\section{Conclusion}
\label{sec:conclusion}

We have presented \texttt{symveig}, a pure NumPy/SciPy package for
rigorous eigenvalue enclosure of Hermitian matrices with an abelian
symmetry-sector decomposition. The enclosures are guaranteed under
IEEE~754 arithmetic by explicit floating-point error analysis, and the
sector decomposition makes them both tighter (by $3$--$9\times$ across
$L = 4$--$12$) and faster (by up to $130\times$ at $L = 12$) than
verifying the full matrix, while never diagonalizing it. We validated the
package on $1$D and $2$D Heisenberg and $J_1$--$J_2$ Heisenberg
Hamiltonians, confirming sound containment in every tested configuration
and a clean, model-independent scaling of the enclosure width with sector
dimension and norm. The package fills a concrete gap: verified
eigenvalue enclosure, previously confined to commercial MATLAB tooling,
made available and symmetry-aware in the scientific-Python stack on which
modern exact diagonalization is built.

\section*{Acknowledgements}


The verified-enclosure methodology builds directly on the work of
S.\ M.\ Rump, M.\ Lange, and S.\ Miyajima.

\bibliographystyle{elsarticle-num}
\bibliography{refs}

\end{document}